\documentclass[preprint]{elsarticle}

\usepackage{graphicx}

\usepackage{amssymb}
\usepackage[tbtags]{amsmath}

\usepackage{bm}
\usepackage{hyperref}
\usepackage{latexsym}

\usepackage{verbatim}

\renewcommand{\d}{ {\rm d} } 
\newcommand{\D}{ {\rm D} } 
\newcommand{\const}{\mathrm{const}}

\newcommand{\iotabar}{\mbox{$\,\iota\!\!$-}}
\newtheorem{remark}{Remark}
\newtheorem{definition}{Definition}

\journal{CNSNS}
\begin{document}
\begin{frontmatter}

\title{Action-gradient-minimizing pseudo-orbits and almost-invariant tori}
\author[a]{R.~L. Dewar\corref{cor1}}
\address[a]{Plasma Research Laboratory, Research School of Physics \& Engineering, The Australian National University, 
ACT~0200, Australia}
\ead{robert.dewar@anu.edu.au}

\author[b]{S.~R. Hudson}
\address[b]{Princeton Plasma Physics Laboratory, PO Box 451, Princeton NJ 08543, USA}
\ead{shudson@pppl.gov}

\author[a]{A.~M. Gibson}

\date{Final, version 1.21 \today}                                           

\begin{abstract}

Transport in near-integrable, but partially chaotic, $1\frac{1}{2}$ degree-of-freedom Hamiltonian systems is blocked by invariant tori and is reduced at \emph{almost}-invariant tori, both associated with the invariant tori of a neighboring integrable system. ``Almost invariant'' tori with rational rotation number can be defined using continuous families of periodic \emph{pseudo-orbits} to foliate the surfaces, while irrational-rotation-number tori can be defined by nesting with sequences of such rational tori. Three definitions of ``pseudo-orbit,'' \emph{action-gradient--minimizing} (AGMin), \emph{quadratic-flux-minimizing} (QFMin) and \emph{ghost} orbits, based on variants of Hamilton's Principle, use different strategies to extremize the action as closely as possible. Equivalent Lagrangian (configuration-space action) and Hamiltonian (phase-space action) formulations, and a new approach to visualizing action-minimizing and minimax orbits based on AGMin pseudo-orbits, are presented.
\end{abstract}

\end{frontmatter}

\section{Introduction}
\label{sec:pseudo}

Periodically forced 1-degree-of-freedom dissipationless physical systems are often called ``$1\frac{1}{2}$ degree-of-freedom (d.o.f.)'' systems \cite{Morrison_98} because their Hamiltonian dynamics is best described in a 3-dimensional phase space (intermediate between the 2-dimensional phase space of a 1-d.o.f. system and the 4-dimensional phase space of a 2-d.o.f. system).

Two important applications of such $1\frac{1}{2}$-d.o.f. systems in plasma physics are to describing one-dimensional single-particle motion in a modulated wave \cite{Elskens_Escande_91,Dewar_Yap_09} and the ``dynamics'' of field lines in nonaxisymmetric toroidal magnetic fields, such as are required in the stellarator approach to fusion plasma confinement \cite{Boozer_04}. An important fluid-mechanics application is to description of passive advection in planar, periodically stirred fluid flows \cite{Metcalfe_10}. 
There are many other applications, so we eschew application-specific notations in the main discussion. However we do, at appropriate points, relate the formalism to the context of the magnetic-field-line dynamics application as this has been the main motivation of this work.

We use standard Hamiltonian dynamics notation \cite{Goldstein_80}, except that we denote the generalized coordinate by $\theta$ rather than $q$, assuming it to be an angle coordinate (in radians so $\theta/2\pi \in S^1$), and we denote its conjugate momentum by $I$, rather than $p$ ($p$ and $q$ being used instead as integers for defining rational fractions).

Also we denote the time-like independent variable by $t$, but it should be borne in mind that in the magnetic plasma confinement application it is not physical time but a toroidal angle \cite{Boozer_04}, and the ``periodic forcing'' is simply the departure of the spatial dependence of the field from axisymmetry. Whether $t$ is time or an angle, we assume $2\pi$-periodicity and endow it with the topology of a circle, $t/2\pi \in S^1$. (However we lift the dynamics to its covering space by taking $t \in \mathbb{R}$ when more convenient.)

The variables $I,\theta$ form polar coordinates in a 2-dimensional phase space, $\Gamma_2$ say. Assuming all motions of interest are bounded in momentum, $I \leq I_{\rm max}$, $\Gamma_2$ is topologically the product of a subset of the real line, $[0,I_{\rm max}]$, and a circle. In $1\frac{1}{2}$-d.o.f. systems it is often convenient to view the dynamical trajectories in an extended \emph{3-dimensional} phase space $\Gamma_3$, by regarding time also as a phase variable. Topologically, as $t$ is angle-like, $\Gamma_3$ is the product of $[0,I_{\rm max}]$ and a 2-torus. (In the magnetic field application $I,\theta,t$ label spatial points in $\mathbb{R}^3$, defining a natural metric in the toroid $\Gamma_3$ embedded in $\mathbb{R}^3$, the dynamical trajectories corresponding to magnetic field lines and invariant tori to magnetic surfaces.)



If a (time-dependent) canonical transformation, $I,\theta \mapsto J,\Theta$, can be found such that the new Hamiltonian is autonomous, then the tori $J = \const$ are invariant under the dynamics and foliate $\Gamma_3$---the system is \emph{integrable}. While $1\frac{1}{2}$-d.o.f. Hamiltonian systems are generically not integrable, they may often be regarded as perturbed away from a neighboring integrable system (not necessarily known in advance). Then chaotic regions in the phase space, arising from homoclinic tangles associated with islands formed at rational rotation number $\omega$, may be separated by residual invariant tori with strongly irrational $\omega$. Transport is completely blocked by invariant tori, but even in chaotic regions it is not uniform \cite{Hudson_Breslau_08}, being slowed at transport barriers associated with ``cantori,'' invariant cantor sets with strongly irrational rotation numbers.

This paper concerns two approaches to defining \emph{almost-invariant} tori joining invariant periodic orbits associated with phase-space islands. (We lose no generality by restricting attention to periodic orbits as sequences of periodic orbits may be used to approximate cantori arbitrarily closely \cite{Meiss_92,Hudson_06a,Hudson_Dewar_09}.)

The general concept of almost-invariant sets acting as transport barriers has been discussed by Froyland and Padberg \cite{Froyland_Padberg_09}, who distinguish \emph{statistical} and \emph{geometrical} approaches to defining almost-invariant sets. The approaches we discuss are geometrical (in $\Gamma_3$) and can be regarded as strategies for minimizing the \emph{action gradient}. 

In discrete-time area-preserving dynamical systems, candidate action-based definitions of ``almost-invariant'' are the quadratic-flux-minimizing (QFMin) topological circles introduced by Meiss and Dewar \cite{Meiss_Dewar_91,Dewar_Meiss_92,Dewar_Khorev_95}, and the ghost circles described in the monograph of Gol\'e \cite{Gole_01}, and references therein. Analogous QFMin and ghost tori were also defined by Hudson and Dewar for Hamiltonian systems \cite{Hudson_Dewar_96} in the context of magnetic field dynamics, but a unified Hamiltonian and Lagrangian formulation has not heretofore been presented.

In this paper we give both Lagrangian (configuration space) and Hamiltonian (phase space) formulations to make clear the similarities and differences of the two viewpoints. In Sec.~\ref{sec:psdyn} we set up the concept of pseudo-orbits and action gradient and in Sec.~\ref{sec:actcont} we develop a new method for visualizing these concepts by reducing the dimensionality of the action extremization problem to 2 through a partial minimization of the action. In Sec.~\ref{sec:AItori} we develop the concept of almost-invariant torus through two different approaches to generating pseudo-orbits, ghost pseudo-orbits and pseudo-orbits that appear naturally as solutions to the Euler--Lagrange equation for the variational problem of minimizing the ``quadratic flux''--- the square of the action gradient integrated over angles.

\section{Dynamics and pseudo-dynamics}
\label{sec:psdyn}

\subsection{Dynamics of $1\frac{1}{2}$ d.o.f. systems}
\label{sec:dynamics}

In terms of the Hamiltonian $H(I,\theta,t)$, the dynamical system in $\Gamma_3$ is
\begin{eqnarray}
	\dot\theta & = &  H_I \;, \label{eq:HamThdot}\\
	        \dot I & = & -H_{\theta} \;, \label{eq:HamIdot}\\
	        \dot t & = & 1 \label{eq:Hamtdot}
\end{eqnarray}
where $H_I$ and $H_{\theta}$ denote the partial derivative of $H(\cdot,\cdot,\cdot)$ with respect to its first and second arguments, respectively, and a dot over a phase variable denotes its total derivative with respect to a dummy time variable, $\tau$ say. 

We assume Eq.~(\ref {eq:HamThdot})  can be solved to give $I$ uniquely as a function of $\theta$, $\dot\theta$ and $t$ so that  the dynamics can be described equivalently by the \emph{Lagrangian}
\begin{equation}
	\label{eq:Ldef}
	L(\theta,\dot\theta,t) \equiv I(\theta,\dot\theta,t)\dot\theta - H(I(\theta,\dot\theta,t),\theta,t) \;,
\end{equation}
in terms of which
\begin{equation}
	\label{eq:Idef}
	I = L_{\dot\theta} \;,
\end{equation}
Eq.~(\ref{eq:HamIdot}) then giving the Lagrangian equation of motion
\begin{equation}
	\label{eq:LEqnMotion}
	\frac{\d}{\d\tau}{L}_{\dot\theta} = L_{\theta} \;,
\end{equation}
where $L_{\theta}$ and $L_{\dot\theta}$ denote the partial derivatives of $L$ with respect to its first and second arguments, respectively.

\begin{remark}[Twist condition]
	\label{rem:TwistNontwist}
A necessary condition for the assumed uniqueness of $I$ as a function of angular velocity $\dot\theta$ is that the \emph{velocity shear} nowhere vanish within the system
\begin{equation}
	\label{eq:ShearCond}
	H_{II} \neq 0 \;.
\end{equation}
If this condition is violated then $I$ and $L$ become multivalued. This case is of interest in both the passive advection \cite{del-Castillo-Negrete_Morrison_93} and magnetic field contexts \cite{Davidson_etal_95}, suggesting that the Hamiltonian approach is more appropriate than the Lagrangian in a wider range of applications.  As the theory of almost invariant tori of the type we discuss in this paper has been predominately developed in a Lagrangian framework we present both formulations and avoid the ``non-twist'' issue by assuming Eq.~\ref{eq:ShearCond}.
\end{remark}


An \emph{invariant} set of phase-space points is one that is mapped onto itself by the dynamics. In the following we consider two kinds of invariant geometric objects lying within $\Gamma_3$: \emph{periodic (closed) paths} and \emph{tori}. 

The special case of an autonomous, 1-d.o.f. Hamiltonian system is \emph{integrable}, meaning action-angle coordinates $J,\Theta$  can be found such that the Hamiltonian, $K$ say, is a function only of $J$. Then the action $J$ is a constant of the motion: $\dot{J} \equiv  -\partial K/\partial\Theta = 0$. The level surfaces of $J$ form a continuous family of invariant 2-tori, nested about a closed invariant loop (periodic orbit) at $J = 0$ and foliating the 3-dimensional $J, \Theta, t$ phase space.  

In action-angle coordinates, $J,\Theta$, the angular velocity is constant on each invariant torus: $\dot\Theta = {\rm const}(J)$. We call $\omega_0(J) \equiv \dot\Theta$ the \emph{rotation number}. (In magnetic plasma confinement it is called the \emph{rotational transform} and denoted by $\iotabar$.) The number-theoretic properties of $\omega_0$ are critically important for understanding the effect of perturbation away from integrability, produced when the Hamiltonian becomes
\begin{equation}
	\label{eq:PertH}
	H = H_0(I) + \epsilon H_1(I,\theta,t) \;,
\end{equation}
to give a non-autonomous $1\frac{1}{2}$-d.o.f. system.

\begin{remark}[Restriction to rotational tori]
\label{rem:diffeo}
Our interest in almost-invar\-iant tori is motivated by a desire \cite{Dewar_85,Hudson_Dewar_99,Hudson_04} to construct global phase-space coordinate systems of an action-angle type using a diffeomorphic canonical transformation from $I,\theta$ to $J,\Theta$ phase-space coordinates. The desired transformation is to be continuously connected to the identity as $\epsilon \to 0$ in such a way that some level sets of $J$ coincide with invariant or almost-invariant tori. Because of this diffeomorphic restriction we restrict attention to \emph{rotational} tori, i.e. tori continuously connected to invariant tori of the unperturbed system, the terminology deriving from an analogy with the physical pendulum. Thus, for our purposes, even integrable perturbations can destroy invariant tori by creating ``islands''  containing invariant tori (librational tori) topologically different distinct from those of the unperturbed system (see Sec.~\ref{sec:actcontPlots}).
\end{remark}

In order that Lagrangian and Hamiltonian approaches be completely equivalent we assume in this paper that Eq.~(\ref{eq:HamThdot}) is always enforced, $\dot\theta -  H_I \equiv 0$, so that Eqs.~(\ref{eq:Ldef}) and (\ref{eq:Idef}) are always valid. We also enforce Eq.~(\ref{eq:Hamtdot}), but to define almost-invariant tori we cannot use exact dynamics so we relax Eq.~(\ref{eq:HamIdot}) as described in detail below.

\subsection{Pseudo-orbits}
\label{sec:psorb}

We start with some definitions:

\begin{definition}[Path]
	\label{def:path}
	A \emph{phase-space path} is a curve in $\Gamma_3$ defined in the $I,\theta,t$ covering space, $[0,I_{\rm max}]\times\mathbb{R}\times\mathbb{R}$, by $\theta = \vartheta(t)$, $I = \mathcal{I}(t)$. A \emph{configuration-space path} is the projection of the curve onto the $\theta,t$ covering space, $\mathbb{R}\times\mathbb{R}$, defined by $\theta = \vartheta(t)$.
\end{definition}
On a path, $\dot\theta \equiv \vartheta'(t)$, $\dot I \equiv \mathcal{I}'(t)$.

\begin{definition} [Periodic path]
	\label{def:periodicpath}
A $(p,q)$-\emph{periodic path}, where $p,q \in \mathbb{Z}$ are mutually prime, is a closed path in $\Gamma_3$ for which the path functions $\vartheta$ and $\mathcal{I}$ obey the \emph{periodicity conditions}
\begin{equation}
\begin{split}
	\label{eq:pqperiodicity}
	\vartheta(t+2\pi q) & =  \vartheta(t) + 2\pi p \;, \\
	\mathcal{I}(t+2\pi q) & =  \mathcal{I}(t) \quad \forall \: t \in \mathbb{R} \;.
\end{split}
\end{equation}
\end{definition}

In the following we assume Eq.~(\ref{eq:Hamtdot}) always to be satisfied, but distinguish different paths by the degree to which they satisfy the dynamical equations of motion:
\begin{definition}[Orbit]
	\label{def:orbit}
An \emph{orbit (periodic orbit)} is a path (periodic path) for which \emph{all} the Hamiltonian equations of motion Eqs.~(\ref{eq:HamThdot}--\ref{eq:Hamtdot}) are exactly satisfied [so, in particular, $\mathcal{I}' + H_{\theta}(\mathcal{I},\vartheta,t) \equiv 0$].
\end{definition}


In this paper we need objects, \emph{pseudo-orbits}, lying between true orbits and arbitrary paths:
\begin{definition}[Pseudo-orbit]
	\label{def:Lpsorb}
A \emph{pseudo-orbit (periodic pseudo-orbit)} is a path (periodic path) on which Eq.~(\ref{eq:HamThdot}) is satisfied exactly, but Eq.~(\ref{eq:HamIdot}) is satisfied only approximately, $\mathcal{I}'  + H_{\theta} = O(\epsilon)$.
\end{definition}

\begin{remark}[Pseudo dynamics]
	\label{rem:PseudoDef}
The term ``pseudo-orbit,'' is used here in the sense introduced by Dewar and Khorev \cite{Dewar_Khorev_95}, which is slightly different from, but in the spirit of, the normal usage in dynamical systems theory \cite{Stewart_Dewar_00,Palmer_09}. Specifically, we do not assume the error (i.e. the norm of the amount by which the dynamical equations fail to be satisfied) is bounded by $\epsilon$, but otherwise arbitrary. Rather, we assume the error is asymptotically $O(\epsilon)$, where $\epsilon$ is a perturbation parameter measuring a departure from integrability. Also, we assume the error terms are well-defined functions of the phase variables, giving a ``pseudo-dynamics,'' rather than arbitrary ``noise.''
	
It should also be realized there is no shadowing theorem in systems close to integrability as they are far from being hyperbolic. On the contrary, a pseudo-orbit may be $O(1)$ away from the closest true orbit.
\end{remark}

\begin{remark}[Hamiltonian/Lagrangian equivalence]
	\label{rem:HamLagEquiv}
As the first condition in Def.~\ref{def:Lpsorb} allows a unique Lagrangian to be defined via Eq.(\ref{eq:Ldef}), the second condition defines a unique Lagrangian pseudo-dynamics $\d L_{\dot\theta}/\d\tau - L_{\theta} = O(\epsilon)$. Conversely, as Eq.~(\ref{eq:Idef}) defines $\mathcal{I}(t)$ uniquely, given $\vartheta(t)$, configuration-space pseudo-orbits map one-to-one to phase-space pseudo-orbits.

In future work it may be useful to allow a more general definition of pseudo-orbit in which Eq.~(\ref{eq:HamThdot}) is only approximately satisfied, but we have imposed $\dot\theta - H_I \equiv 0$ in this paper so that the Hamiltonian pseudo-dynamics we develop here is equivalent to the Lagrangian pseudo-dynamics we have previously used.
\end{remark}

\subsection{Action gradients}
\label{sec:actgrad}

The \emph{action} of a $(p,q)$-periodic configuration-space path $\theta = \vartheta(t)$ is defined as a functional of $\vartheta$ by the integral
\begin{equation}
	S[\vartheta]=\int^{2\pi q}_0 L(\vartheta,\vartheta',t)\, \d t \;.
	\label{eq:actiondef}
\end{equation}
Correspondingly, the \emph{phase-space action} of a $(p,q)$-periodic phase-space path  $\theta = \vartheta(t)$, $I = \mathcal{I}(t)$ is defined by
\begin{equation}
	S_{\rm ph}[\vartheta,\mathcal{I}]
	=\int^{2\pi q}_0 [\mathcal{I}\vartheta' - H(\mathcal{I},\vartheta,t)]\, \d t \;.
	\label{eq:phactiondef}
\end{equation}

The first variations are linear functionals of the variations, $\delta\vartheta(t)$ and $\delta\mathcal{I}(t)$, of the path functions,
\begin{equation}
	\delta S = \left\langle \delta\vartheta, \frac{\delta S}{\delta\theta} \right\rangle \;,
	\label{eq:actionvar}
\end{equation}
and
\begin{eqnarray}
	\delta S_{\rm ph} 
	& = & \left\langle \delta\vartheta, \frac{\delta S_{\rm ph}}{\delta\theta} \right\rangle 
	+ \left\langle \delta\mathcal{I}, \frac{\delta S_{\rm ph}}{\delta I} \right\rangle \nonumber \\
	& \equiv & \left\langle
	  [\delta\vartheta, \delta\mathcal{I}],
	  \left[\frac{\delta S_{\rm ph}}{\delta\theta},
	   \frac{\delta S_{\rm ph}}{\delta I}\right]^{\rm T} 
	\right\rangle\;,
	\label{eq:phactionvar}
\end{eqnarray}
where $\langle f, g\rangle$ denotes the $L^2$ inner product between arbitrary path functions $f$ and $g$,
\begin{equation}
	\langle f,g\rangle \equiv \int_0^{2\pi q} fg \, \d t \;,
	\label{eq:innerprod}
\end{equation}
and $^{\rm T}$ denotes matrix transpose.
Thus $\delta S/\delta\theta$ and $[\delta S_{\rm ph}/\delta\theta, \delta S_{\rm ph}/\delta I ]^{\rm T}$ may be regarded as infinite-dimensional action gradients.

Varying $\vartheta$ and $\mathcal{I}$ in Eqs.~(\ref {eq:actiondef}) and (\ref {eq:phactiondef}), integrating by parts and comparing with Eqs.~(\ref {eq:actionvar}) and (\ref {eq:phactionvar}) we make the identifications
\begin{equation}
	\frac{\delta S}{\delta\theta} = L_{\theta} - \frac{\d L_{\dot\theta}}{\d t} \;,
	\label{eq:actiongrad}
\end{equation}
and
\begin{eqnarray}
	\frac{\delta S_{\rm ph}}{\delta\theta} & = & -\dot I - H_{\theta} \;.
	\label{eq:phactiongradtheta}\\
	\frac{\delta S_{\rm ph}}{\delta I} & = & \dot\theta - H_I \;.
	\label{eq:phactiongradI}
\end{eqnarray}
Thus, comparing the above with Def.~\ref{def:orbit}, we see that \emph{the action gradient vanishes on a true physical orbit}. This is a statement of Hamilton's Principle---orbits extremize the action: $\delta S = 0$ $\forall \,\delta\vartheta,\delta\mathcal{I}$.

Definition~\ref{def:Lpsorb} requires a pseudo-orbit to satisfy Eq.~(\ref{eq:HamThdot}) exactly, which, comparing with Eq.~(\ref{eq:phactiongradtheta}) is the constraint on the phase-space action gradient
\begin{equation}
	\frac{\delta S_{\rm ph}}{\delta I} = 0 \;.
	\label{eq:AgradHamThdot}
\end{equation}

Therefore, $[\delta S_{\rm ph}/\delta\theta,$ $\delta S_{\rm ph}/\delta I ] \equiv [\delta S/\delta\theta, 0]$ on a pseudo-orbit.  Consequently, whether working in phase space or configuration space, we shall mean by the unqualified term \emph{action gradient} the quantity $\delta S/\delta\theta$. 

\section{AGMin pseudo-orbits and action contours}
\label{sec:actcont}

The space of all possible paths is infinite-dimensional and thus the function $S$ and its gradient $\delta S/\delta\theta$ are difficult to visualize. Although, as described in Sec.~\ref{sec:ghost}, the gradient flow induced on a periodic orbit by $\delta S/\delta\theta$ forms a ghost torus in $\Gamma_3$, this only gives a partial picture of the nature of $S$. Instead, in this section we present a method for selecting a two-parameter family of $(p,q)$-periodic pseudo-orbits that includes the true $(p,q)$-periodic orbits at the minima and minimax points of $S$ defined on this family.

\begin{figure}[htbp]
   \centering
       \begin{tabular}{cc}
		\includegraphics[width = 5.5cm]{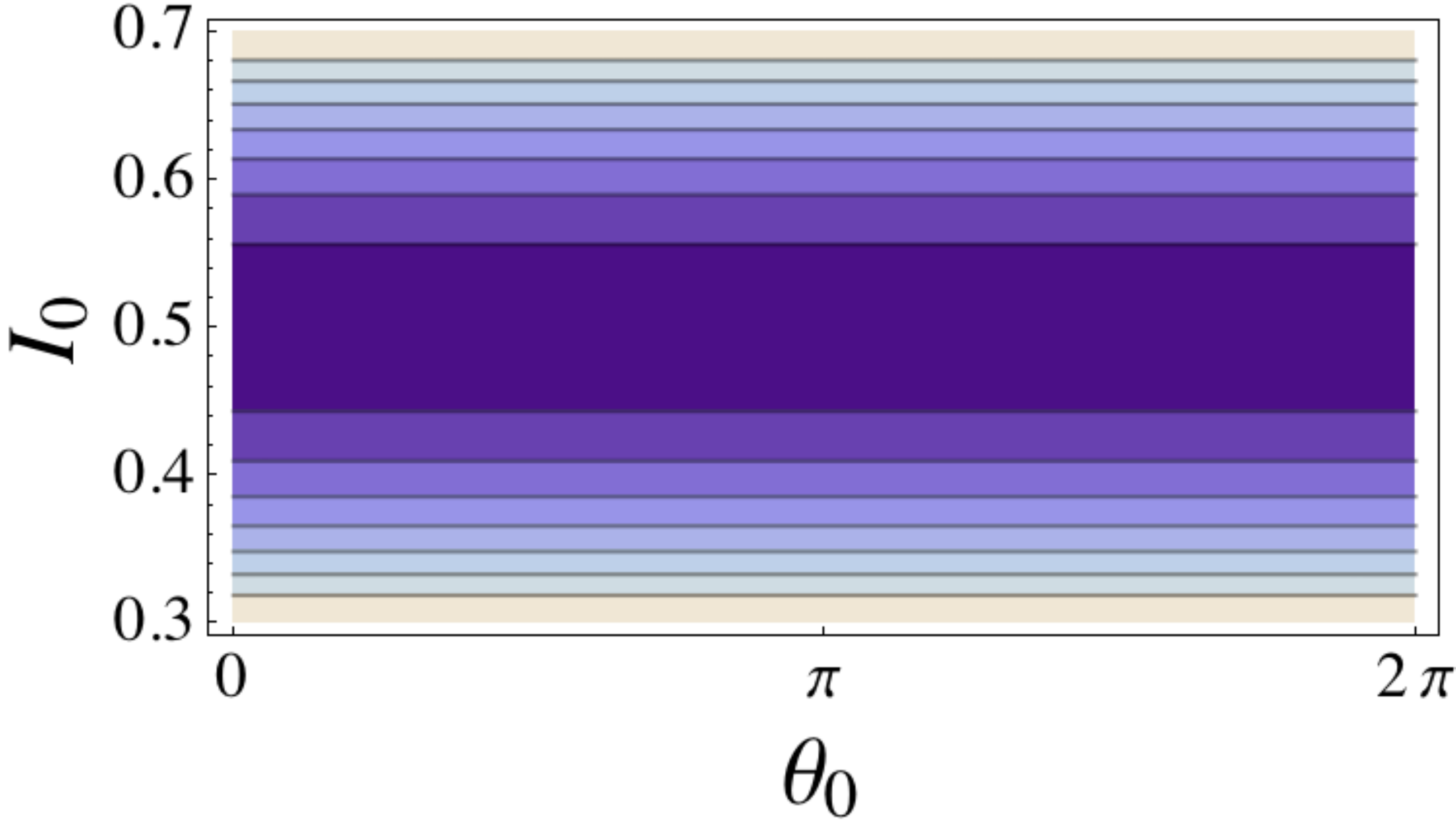} 
	&
		\includegraphics[width = 5.5cm]{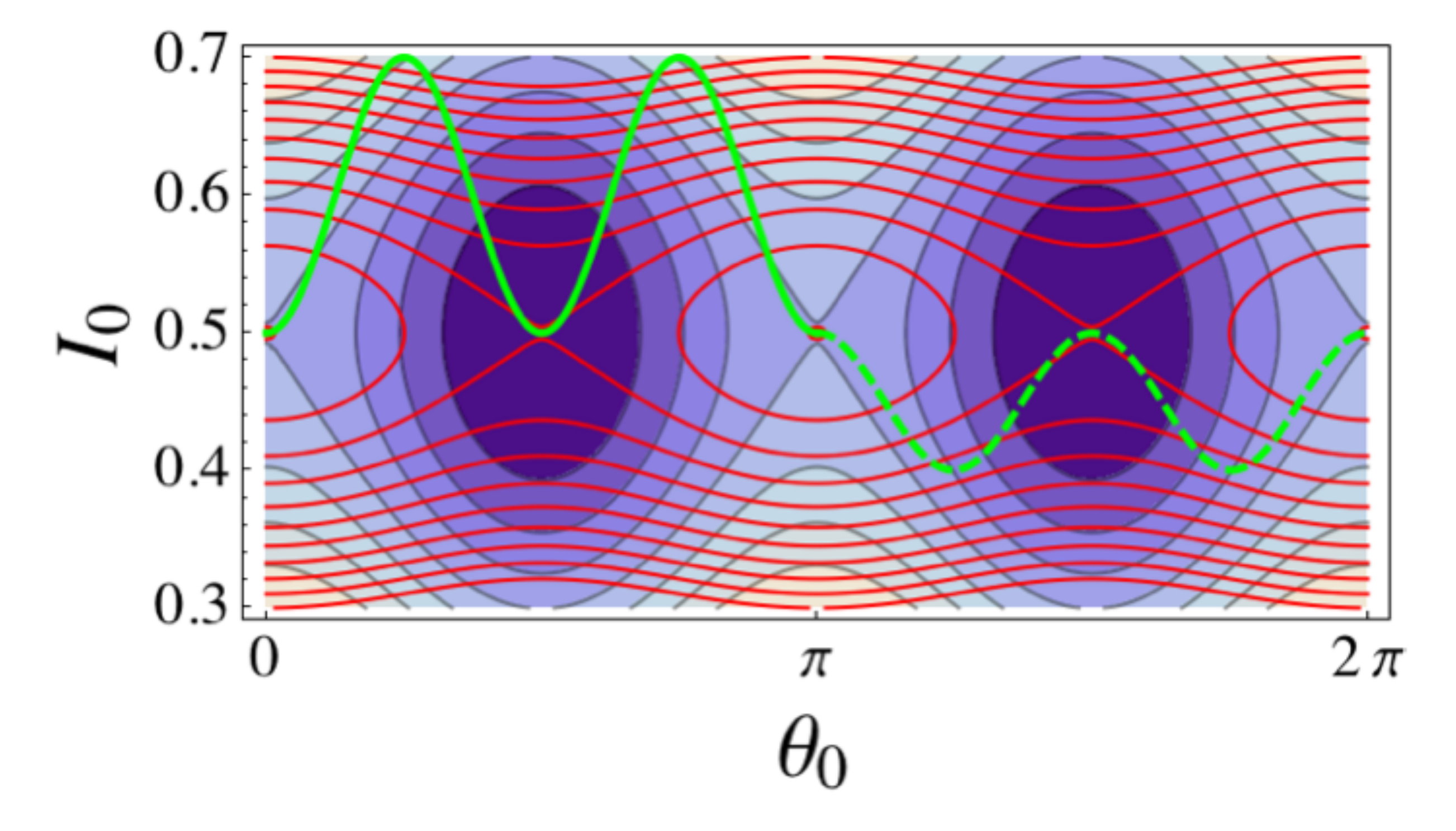}
	\end{tabular}
   \caption{Action contours (color online) for AGMin $(1,2)$-periodic pseudo-orbits described in Sec.~(\ref{sec:actcontPlots}).  (a) Left:  the unperturbed case $\epsilon = 0$, in which there is a line of $(1,2)$-periodic orbits at $I_0 = 1/2$. (b) Right: perturbed case $\epsilon = 0.002$, in which all but two $(1,2)$-periodic orbits, the action-minimizing and -minimax orbits, are destroyed. Also shown in (b) are rest-frame energy contours (red solid curves) to illustrate the fact that minimax/minimum points of action are minimum/minimax points of energy, respectively, and the set of starting points (solid green curve) used to illustrate the pseudo-invariant torus concept in Sec.~\ref{sec:pstor}, and its image (dashed green curve) under the return map.}
   \label{fig:SInteg}
\end{figure}

The method is based on selecting only pseudo-orbits that minimize the action gradient along a path in the $L^2$ norm, $\|\delta S/\delta\theta\| \equiv \langle \delta S/\delta\theta,\delta S/\delta\theta\rangle^{1/2} \geq 0$, where the inner product is defined in Eq.~(\ref{eq:innerprod}).
\begin{definition}[AGMin pseudo-orbits]
	\label{def:QFMinOrb}
	
\emph{Action-gradient-minimizing (AGMin) pseudo-orbits are} paths that minimize the action-gradient norm $\|\delta S/\delta\theta\|$, which vanishes if and only if $\delta S/\delta\theta \equiv 0$ on the entire path, i.e. on true orbits. 
\end{definition}

\subsection{Lagrangian variational principle for AGMin pseudo-orbits}
\label{sec:Lagactcont}

A $(p,q)$-periodic AGMin pseudo-orbit minimizes the \emph{AGMin objective functional}
\begin{equation}
	f_{p,q}[\vartheta] \equiv  \frac{1}{2}\int_0^{2\pi q} \left(\frac{\delta S}{\delta\theta}\right)^2 \, \d t
	\label{eq:POrbQF}
\end{equation}
over $(p,q)$-periodic paths $\vartheta_{p,q}(t)$.

From Eq.~(\ref{eq:actiongrad}) 
\begin{eqnarray}
	\delta\frac{\delta S}{\delta\theta}
	& =  &L_{\theta\theta}\delta\vartheta + L_{\theta\dot\theta}\delta\vartheta' 
	- \frac{\d} {\d t}\left(L_{\theta\dot\theta}\delta\vartheta + L_{\dot\theta\dot\theta}\delta\vartheta' \right)
	\nonumber \\
	& =  &\left(L_{\theta\theta} - \frac{\d L_{\theta\dot\theta}}{\d t}\right)\delta\vartheta
	 +\frac{\d} {\d t}\left[\left(\frac{\d L_{\dot\theta\dot\theta}}{\d t}\right)\delta\vartheta
	- \frac{\d} {\d t}\left(L_{\dot\theta\dot\theta}\delta\vartheta\right)
	\right] \;.
	\label{eq:POrbdeltaactiongrad}
\end{eqnarray}
Substituting Eq.~(\ref{eq:POrbdeltaactiongrad}) and varying Eq.~(\ref{eq:POrbQF}) we find, after integrating by parts,
\begin{equation}
	\delta f_{p,q}  =  \int_0^{2\pi q} \!\!\!\!\delta\vartheta
	\left[\left(L_{\theta\theta} - \frac{\d L_{\theta\dot\theta}}{\d t}\right)
			- \frac{\d} {\d t}L_{\dot\theta\dot\theta}\frac{\d} {\d t}
	\right]
	\frac{\delta S}{\delta\theta} \, \d t \;.
	\label{eq:POrbQFvar}
\end{equation}
Setting $\delta f_{p,q} = 0$ $\forall\,\delta\vartheta$ we find the Euler--Lagrange equation for AGMin pseudo-orbits
\begin{equation}
	\left[\left(L_{\theta\theta} - \frac{\d L_{\theta\dot\theta}}{\d t}\right)
			- \frac{\d} {\d t}L_{\dot\theta\dot\theta}\frac{\d} {\d t}
	\right]
	\frac{\delta S}{\delta\theta} = 0 \;,\; \forall\, t\in [0,2\pi q] \;,
	\label{eq:POrbQFEL}
\end{equation}
where $\delta S/\delta\theta$ stands for the expression in Eq.~(\ref{eq:actiongrad}).
In Eqs.~(\ref{eq:POrbQFvar}) and (\ref{eq:POrbQFEL}) the second-order time derivative term has been written in such a way that its self-adjointness is manifest, $\d/\d t$ being regarded as an operator that acts on everything to its right except when its scope is limited by being inside parentheses $(\ldots)$. 

From Eq.~(\ref{eq:actiongrad}) we see that Eq.~(\ref{eq:POrbQFEL}) is a fourth-order differential equation for $\vartheta$. The four arbitrary constants in its general solution are to be determined from two initial conditions and two periodicity conditions (cf. Def.~\ref{def:periodicpath})
\begin{eqnarray}
	\vartheta(0) & = & \theta_0 \;, \nonumber\\
	\vartheta'(0) & = & \dot\theta_0 \;, \nonumber\\
	\vartheta(2\pi q) & = & \theta_0 + 2\pi p \;, \nonumber\\
	\vartheta'(2\pi q) & = & \dot\theta_0  \;.
	\label{eq:pqperiodicityconds}
\end{eqnarray}

These boundary conditions ensure that the periodic extension to all $t$ of the pseudo-orbit segment obtained by solving Eq.~(\ref{eq:POrbQFEL}) satisfies continuity of $\vartheta(t)$ and $\vartheta'(t)$ $\forall\, t \in \mathbb{R}$. However, for arbitrary $\dot\theta_0$, $\vartheta''(t)$ will in general be discontinuous at $t = 2\pi qk$, $k\in \mathbb{Z}$.

\subsection{Action contours}
\label{sec:actcontPlots}

In Figure~\ref{fig:SInteg} we show two action-contour plots using the particle-in-wave Lagrangian/Hamiltonian
\begin{equation}
	\label{eq:pendulum}
	L = \frac{{\dot\theta}^2}{2}+\epsilon\cos(m \theta - n t)\;,\: H = \frac{I^2}{2}-\epsilon\cos(m \theta - n t) \;,
\end{equation}
taking, specifically, $m=2$, $n=1$ (which also models a magnetic island at the $\iotabar = 0.5$ surface). The contours of the action were plotted for the 2-parameter family of $(1,2)$-periodic AGMin pseudo-orbits: Setting $p =1$, $q = 2$, Eq.~(\ref{eq:POrbQFEL}) under the boundary conditions Eqs.~(\ref{eq:pqperiodicityconds}) was solved (using the \emph{Mathematica} \cite{Mathematica8} routine \texttt{NDSolve}), for an array of initial conditions $\theta_0$ and $\dot\theta_0 = I_0$ and plotted using \emph{Mathematica}'s \texttt{ListContourPlot}.

Figure~\ref{fig:SInteg}(a) shows the unperturbed case, $\epsilon = 0$. In this case there is a horizontal valley of minima of $S$ on the rational invariant torus $I_0 = \omega = 1/2$, which is foliated by a family of $(1,2)$-periodic orbits (as $\delta S = 0$ for each member of this family, all members must have the same value  of $S$).

Figure~\ref{fig:SInteg}(b) shows the case $\epsilon = 0.002$, where all but two of the $\omega = 1/2$ periodic orbits are destroyed because the action valley is not structurally stable under perturbation: the slightest ripple breaks it into minima and saddle points. Only two periodic orbits survive: the ``minimax'' orbit passing through the saddle points of $S$, which are seen to coincide with wave-frame-energy minima and hence this orbit is elliptically stable; and the ``minimizing'' orbits passing through the minima of $S$, which coincide with the  saddle points of wave-frame energy (red curves, see below) and are thus hyperbolically unstable. The green curves are explained in Sec.~\ref{sec:pstor}.

By making a Galilean transformation to the wave frame the system in Eq.~(\ref{eq:pendulum}) can be made autonomous (it is isomorphic to the physical pendulum), and is thus integrable. The red solid curves in Fig.~\ref{fig:SInteg}(b) show energy contours in the wave frame and are thus $t=0$ sections of invariant tori, which are seen to be of two distinct topological types---``librating'' orbits within the island separatrix, and ``rotating'' orbits outside the separatrix. There is no diffeomorphic action-angle transformation continuously connected to the identity so this example, though integrable, is sufficient for illustrating the destruction of a primary invariant torus (see Remark~\ref{rem:diffeo}), specifically the torus with the resonant rotation number $\omega = n/m = 0.5$. [Note that, unlike the $(1,2)$-periodic pseudo-orbits used to construct the action-contour plots, each true orbit in general has a different rotation number.]

The path-pseudo-orbit method thus provides a visualization of the Poincar\'e--Birkhoff theorem \cite{Meiss_92}, which shows that survival of a pair of minimizing and minimax orbits is generic after invariant torus with rational rotation number of an integrable system is destroyed by a perturbation that (in general) makes the system non-integrable. The minimizing and minimax orbits thus make a robust framework on which to build a theory of almost-invariant tori, as shown in Sec.~\ref{sec:AItori}.

\begin{figure}[htbp]
   \centering
       \begin{tabular}{cc}
		\includegraphics[height = 5.5cm]{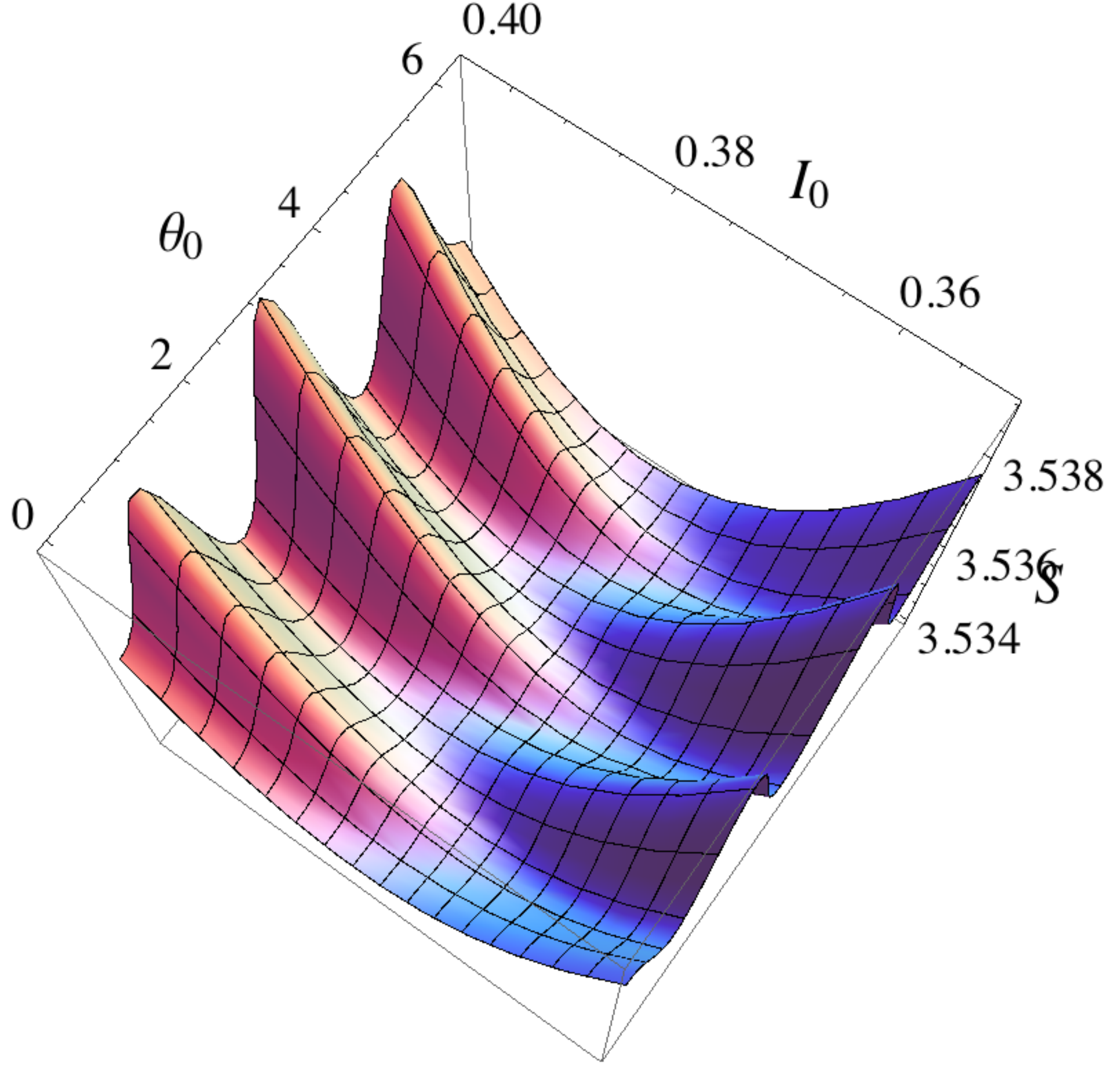} 
	&
		\includegraphics[height = 5cm]{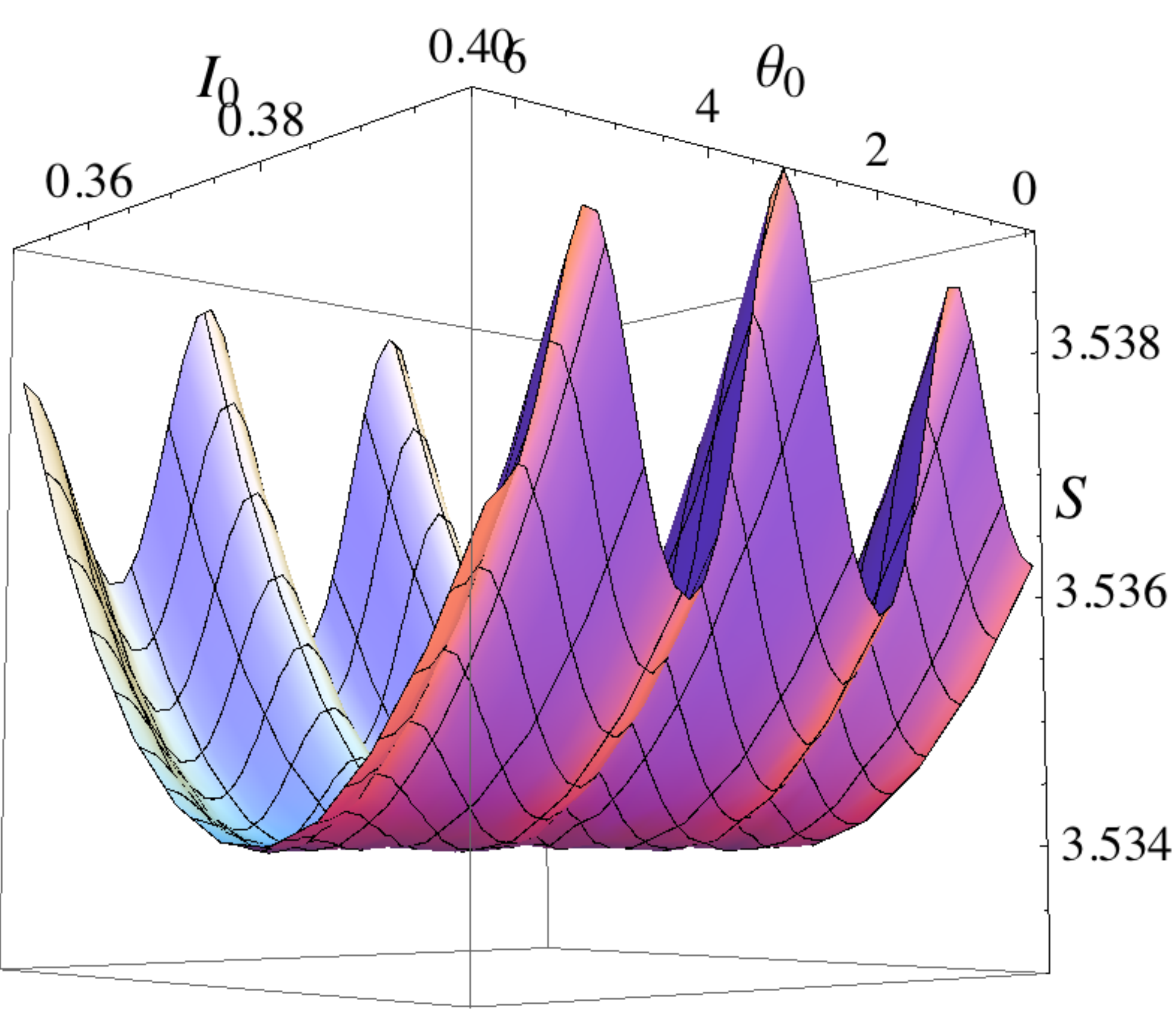}
	\end{tabular}
   \caption{Three-dimensional plots (color online) of action $S$ \emph{vs.} $I_0,\theta_0$ for AGMin $(3,8)$-periodic pseudo-orbits, computed in the same way as the pseudo-orbits in Fig.~\ref{fig:SInteg}, for the nonintegrable Hamiltonian defined in Eq.~(\ref{eq:2wave}) and following text. (a) Left:  Top view showing the nonlinearly deformed ``action valley'' corresponding to the almost-invariant torus with $\omega = 3/8$. (b) Right: Bottom view showing that the valley floor is almost level.}
   \label{fig:S38}
\end{figure}

\subsection{Hamiltonian variational principle for AGMin pseudo-orbits}
\label{sec:Hamactcont}

In the alternative Hamiltonian phase-space formulation we introduce a Lagrange multiplier $\lambda(t)$ to take into account the constraint imposed by Eq.~(\ref{eq:AgradHamThdot}), which is difficult to implement explicitly. Then, to define $(p,q)$-periodic AGMin pseudo-orbits we extremize the functional
\begin{equation}
	f_{{\rm ph}\, p,q} [\mathcal{I},\vartheta,\lambda] \equiv
	 \int_0^{2\pi q} \left[\frac{1}{2}\left(\frac{\delta S_{\rm ph}}{\delta\theta}\right)^2
	- \lambda\frac{\delta S_{\rm ph}}{\delta I} \right] \d t
	\label{eq:POrbphQF}
\end{equation}
under independent variations of $\mathcal{I}$ and $\vartheta$.
Varying Eqs.~(\ref{eq:phactiongradtheta}) and (\ref{eq:phactiongradI})
\begin{eqnarray}
	\delta\frac{\delta S_{\rm ph}}{\delta\theta}
	& = &  -\delta{\mathcal{I}}' - H_{I\theta}\delta\mathcal{I} - H_{\theta\theta}\delta\vartheta  \;,
	\label{eq:POrbactiongradthetavar}\\
	\delta\frac{\delta S_{\rm ph}}{\delta I} & = & \delta\vartheta' - H_{II}\delta\mathcal{I} - H_{I\theta}\delta\vartheta  \;,
	\label{eq:POrbphQFSvar}
\end{eqnarray}
where ${\mathcal{I}}'$ denotes $\mathcal{I}'(t)$.  Using these expressions and varying Eq~(\ref{eq:POrbphQF}) we find
\begin{eqnarray}
	\delta f_{{\rm ph}\, p,q} & = &
	\!\!\int_0^{2\pi q} \!\!\!\!\d t\left\{
	\delta\mathcal{I}
		\left[
			\left(\frac{\d} {\d t} - H_{I\theta}\right)\frac{\delta S_{\rm ph}}{\delta\theta} + \lambda H_{II}
		\right] \right.\nonumber\\
	&&\left.\quad\mbox{} + \delta\vartheta
	\left[-H_{\theta\theta}\frac{\delta S_{\rm ph}}{\delta\theta}
		+ \left(\frac{\d} {\d t} + H_{I\theta}\right)\lambda
	\right]
	\right\} \;.
	\label{eq:POrbphQFvar}
\end{eqnarray}
Requiring $\delta f_{\rm ph} = 0$ $\forall\,\delta\mathcal{I},\,\delta\vartheta$ gives the two Euler--Lagrange equations
\begin{eqnarray}
	\left(\frac{\d} {\d t} - H_{I\theta}\right)\frac{\delta S_{\rm ph}}{\delta\theta} + \lambda H_{II} & = & 0
	\label{eq:POrbphQFEL1} \;,\\
	-H_{\theta\theta}\frac{\delta S_{\rm ph}}{\delta\theta}
		+ \left(\frac{\d} {\d t} + H_{I\theta}\right)\lambda & = & 0 \;,
	\label{eq:POrbphQFEL2}
\end{eqnarray}
where $\delta S_{\rm ph}/\delta\theta$ stands for the expression in Eq.~(\ref{eq:phactiongradtheta}).

These comprise one second-order and one first-order ordinary differential equation. With the constraint Eq.~(\ref{eq:AgradHamThdot}), which is a first-order o.d.e., we have three equations for the three dependent variables, $\vartheta$, $\cal I$ and $\lambda$. As we assume the twist condition Eq.~(\ref{eq:ShearCond}) we can eliminate $\lambda$ between Eqs.~(\ref{eq:POrbphQFEL1}) and Eq.~(\ref {eq:POrbphQFEL2}) and replace them with the single third-order o.d.e.
\begin{equation}
		\left[
			\left(\frac{\d} {\d t} + H_{I\theta}\right)\frac{1}{H_{II}}\left(\frac{\d} {\d t} - H_{I\theta}\right)
			+ H_{\theta\theta}
		\right]\frac{\delta S_{\rm ph}}{\delta\theta} =  0 \;.
	\label{eq:POrbphQFEL3}
\end{equation}

The four arbitrary constants of the general solution are to be determined from the initial conditions and periodicity,
\begin{eqnarray}
	\mathcal{I}(0) & = & I_0 \;, \nonumber\\
	\vartheta(0) & = & \theta_0 \;, \nonumber\\
	\mathcal{I}(2\pi q) & = & I_0 \;, \nonumber\\
	\vartheta(2\pi q) & = & \theta_0 + 2\pi p\;.
	\label{eq:phperiodicityconds}
\end{eqnarray}

It is readily verified that the Hamiltonian pseudo-orbit defining equations Eq.~(\ref{eq:AgradHamThdot}), Eq.~(\ref{eq:POrbphQFEL3}) and Eqs.~(\ref{eq:phperiodicityconds}) are equivalent to the Lagrangian defining equations Eq.~(\ref{eq:POrbQFEL}) and Eqs.~(\ref{eq:pqperiodicityconds}).

Figure~\ref{fig:S38} shows action plots for a particle in two waves (or field lines affected by islands on two rational surfaces), which is a nonintegrable system with Hamiltonian
\begin{equation}
	\label{eq:2wave}
	H = \frac{I^2}{2} + \epsilon V(\theta,t) \;,
\end{equation}
where $V(\theta,t) \equiv -(1-\delta)\cos(m_1 \theta - n t) - \delta\cos(m_2 \theta - n t)$. 
Specifically, we take $n=1$, $m_1 = 2$ and $m_2 = 3$, which excites resonant islands at $\omega = 1/2$ and $1/3$, and take $\delta = 0.5$, $\epsilon = 4\times 10^{-4}$. In the boundary conditions Eq.~(\ref{eq:pqperiodicityconds}) or Eq.~(\ref{eq:phperiodicityconds}) we take $p=3$, $q=8$ so the path-pseudo-orbits have rotation number $\omega = 3/8 = 0.375$. This fraction is a close rational approximant to the noble number $1-\gamma^{-1} = 0.381966\ldots$, where $\gamma \equiv (1+\sqrt 5)/2$ is the golden mean (cf. Fig.~6 of Ref.~\citenum{Dewar_Meiss_92}). The AGMin pseudo-orbit defining equations were again solved using \emph{Mathematica}'s \texttt{NDSolve}, but in Fig.~\ref{fig:S38} the results are displayed using \texttt{ListPlot3D}.

As the invariant torus at $\omega = 1-\gamma^{-1}$ is very robust it should survive up to $\epsilon = O(1)$ so the pseudo-orbits of the neighboring $\omega = 3/8$ almost-invariant-torus should all satisfy $\delta S \approx 0$ for the small value of $\epsilon$ we use. Thus the floor of the action valley should be almost level, as is observed, illustrating the potential of AGMin pseudo-orbits as a visualization tool for illustrating the Kolmogorov--Arnold--Moser (KAM) theorem \cite{Meiss_92,Arrowsmith_Place_91} and testing the existence of KAM tori.

\section{Almost-invariant tori}
\label{sec:AItori}

\begin{figure}[htbp]
   \centering
       \begin{tabular}{cc}
		\includegraphics[height = 5cm]{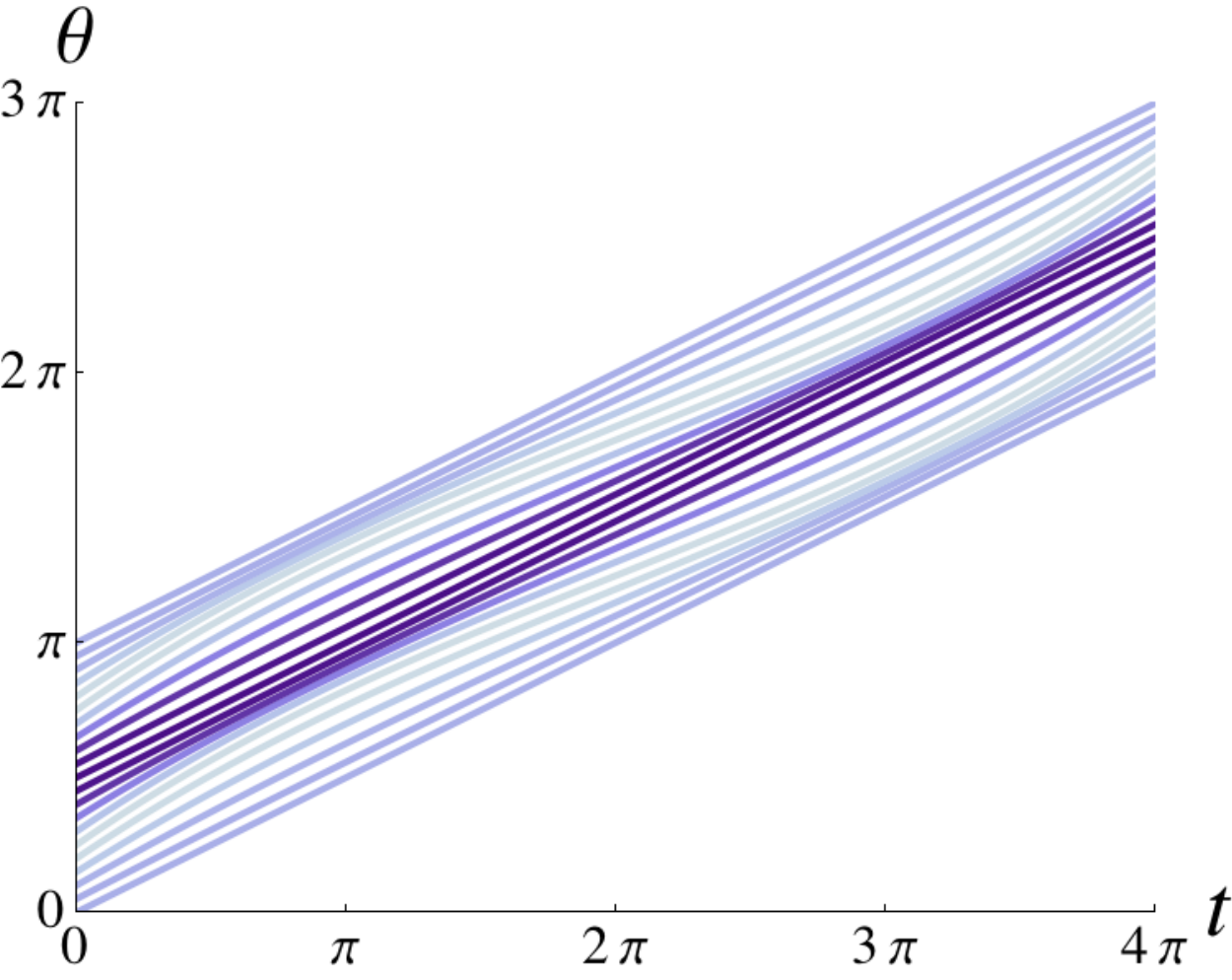} 
	&
		\includegraphics[height = 5cm]{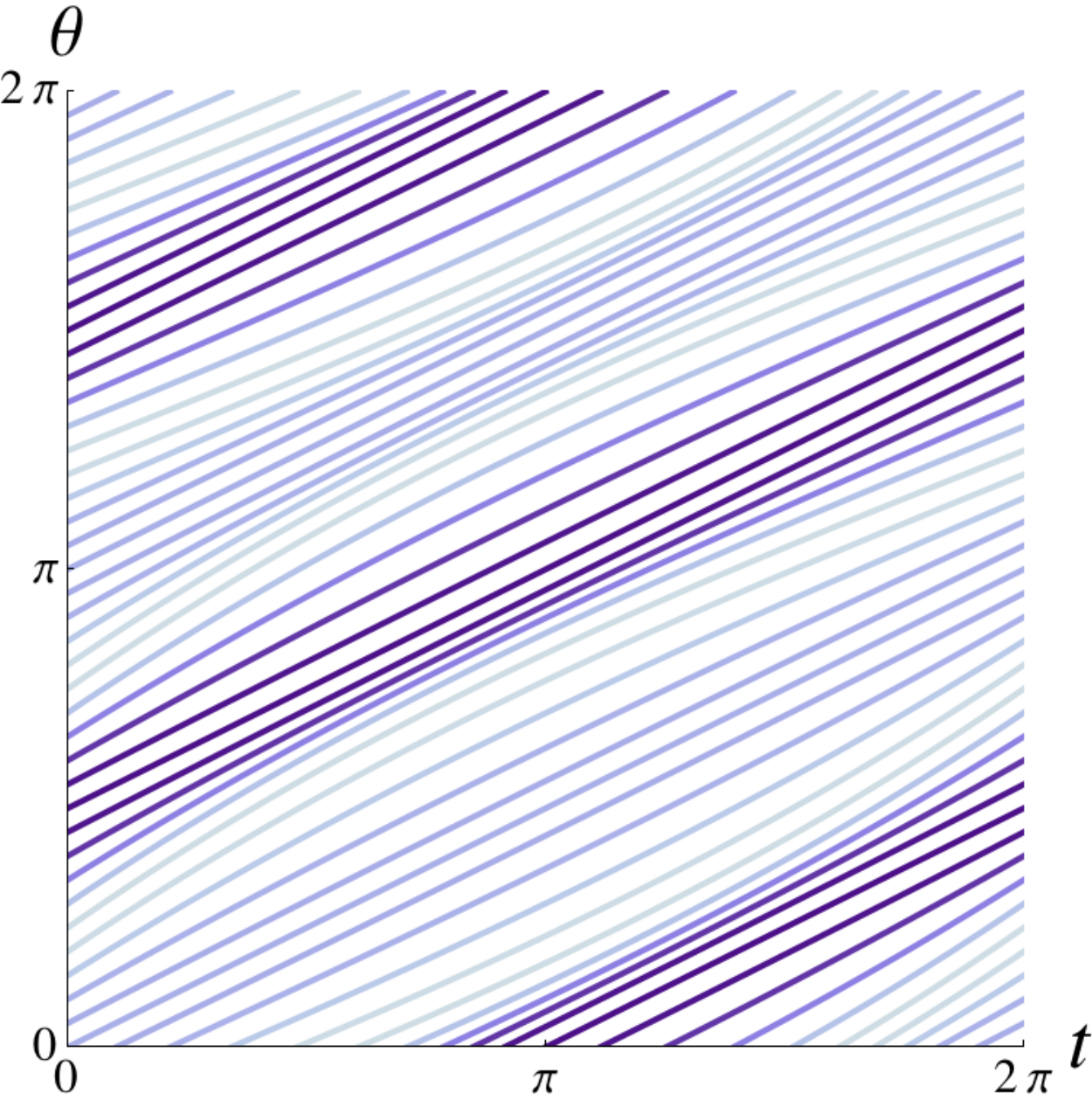}
	\end{tabular}
   \caption{Illustration of the covering of a torus using a subset of the $\epsilon = 0.002$ AGMin (1,2)-periodic pseudo-orbits described in Sec.~\ref{sec:actcont} and plotted using \emph{Mathematica}'s \texttt{ParametricPlot}. (a) Left: pseudo-orbit set described below plotted in the $\theta,t$ covering space. (b) Right: the same set applying torus topology.}
   \label{fig:psTorusConfig}
\end{figure}

\subsection{Pseudo-invariant tori}
\label{sec:pstor}

To generalize the concept of invariant torus, we consider a 1-parameter family of $(p,q)$-periodic pseudo-orbits $\vartheta(t|\theta_0)$, $2\pi$-periodic in $\theta_0$, such that $\vartheta(0|\theta_0) \equiv \theta_0$. The almost-invariant tori studied in this paper are specifically those that can be built from from these pseudo-orbits. 

However, before giving the full construction we first single out the pseudo-orbit starting at $\theta_0 = 0$ and consider the set formed from its images under the return map: $\{\vartheta(2\pi j|0) \!\!\mod 2\pi|j = 0,1,\ldots, q-1\}$. We then rearrange this set to form the ordered set $\{\theta_0^{(0)}\equiv 0, \theta_0^{(1)}, \ldots, \theta_0^{(q-1)}\}$, $\theta_0^{(i)} < \theta_0^{(i+1)}$ and extend it by defining $\theta_0^{(q)} \equiv 2\pi$ [as $\vartheta(2\pi q|0)$ and $\vartheta(0|0)$ are topologically equivalent by Eq.~(\ref{eq:pqperiodicity})].

Recalling that Def.~\ref{def:Lpsorb} requires Eq.~(\ref{eq:HamThdot}) to be satisfied, we can use Eq.~(\ref{eq:Idef}) to determine $I$ and define a
\begin{definition}[Pseudo-invariant torus]
	\label{def:psitorus}
A \emph{pseudo-invariant torus} $\mathcal{T}_{p,q}[\vartheta]$  is a surface swept out in $\Gamma_3$ by a family of $(p,q)$-periodic pseudo-orbits 
\begin{eqnarray}
	\label{eq:torusfoliation}
	t & = & \tau \mod 2\pi \;, \nonumber \\
	\theta & = & \vartheta(\tau|\theta_0) \mod 2\pi \;, \nonumber \\
	        I & = & L_{\dot\theta}(\vartheta(\tau|\theta_0),\d_{\tau}\vartheta(\tau|\theta_0),\tau) \quad \forall \: \tau \in [0,2\pi q)
\end{eqnarray}
as $\theta_0$ varies over the range $$\theta_0 \in [0,\theta_0^{(1)})\;,$$ where $\theta_0^{(1)}$ is the first nonzero member of the ordered set of return points defined above.
\end{definition}

In the above $\d_{\tau} \equiv \partial/\partial \tau$ denotes the time derivative along a given pseudo-orbit, i.e. with $\theta_0$ fixed. By restricting the range of $\theta_0$ we create a helical ribbon which, after wrapping around the torus $q$ times covers it completely. This is illustrated in Fig.~\ref{fig:psTorusConfig} using a set of (1,2)-periodic AGMin-pseudo-orbits having the set of initial values shown by the solid green sinusoidal curve in Fig.~\ref{fig:SInteg}(b). The dashed green sinusoidal curve in Fig.~\ref{fig:SInteg}(b) shows the image of the solid curve under the return map---taken together these two green curves cover the interval $\theta_0 \in [0,2\pi]$. 

Note that the set of pseudo-orbits has been selected to include the minimizing and minimax true orbits as this is a natural feature of an almost-invariant torus (see Secs.~\ref{sec:ghost} and \ref{sec:QFMinTori}). However this is not an essential feature of a pseudo-invariant torus, nor is the use of AGMin pseudo-orbits.

\begin{figure}[htbp]
   \centering
       \begin{tabular}{cc}
		\includegraphics[height = 4.75cm]{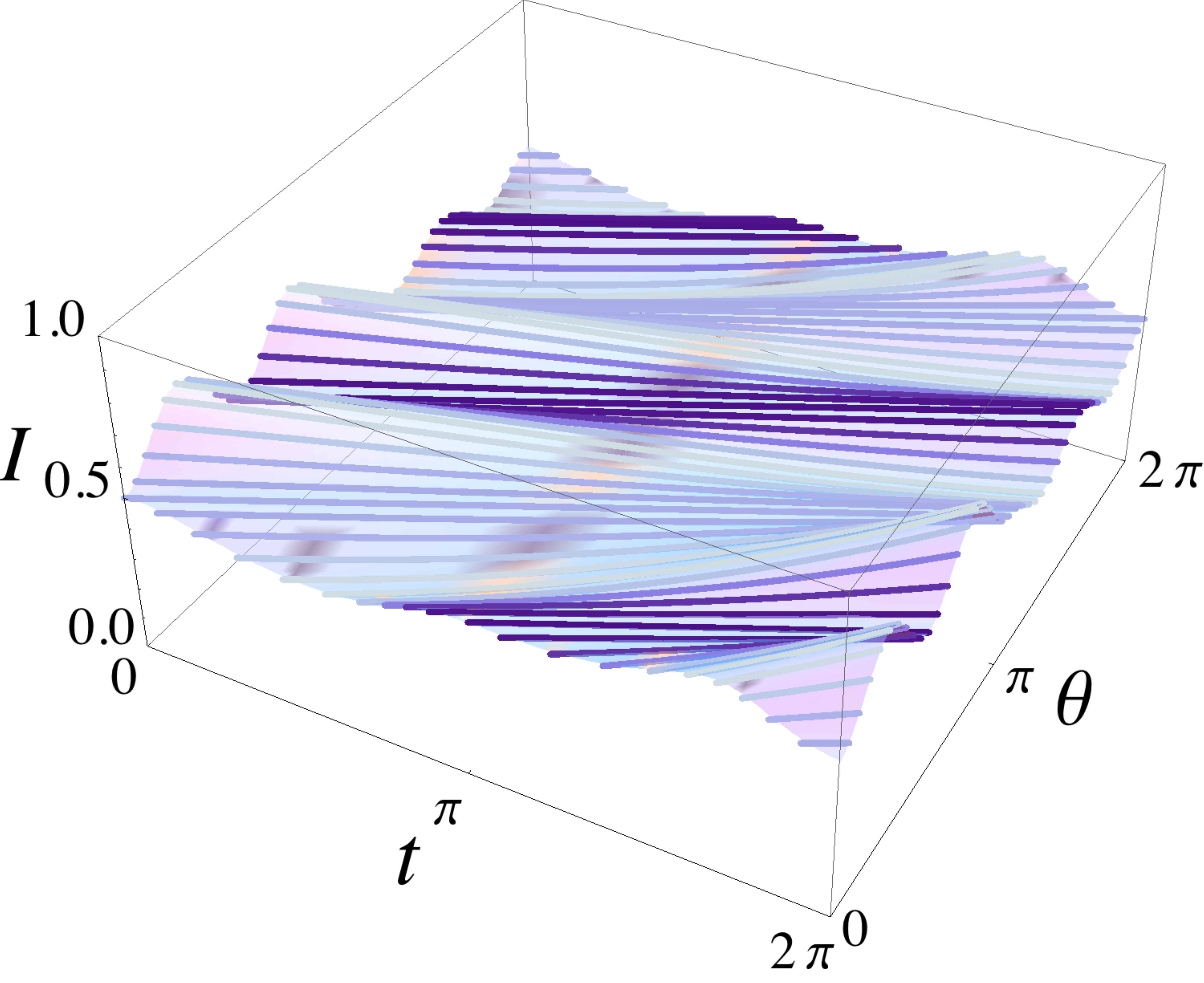} 
	&
		\includegraphics[height = 4.75cm]{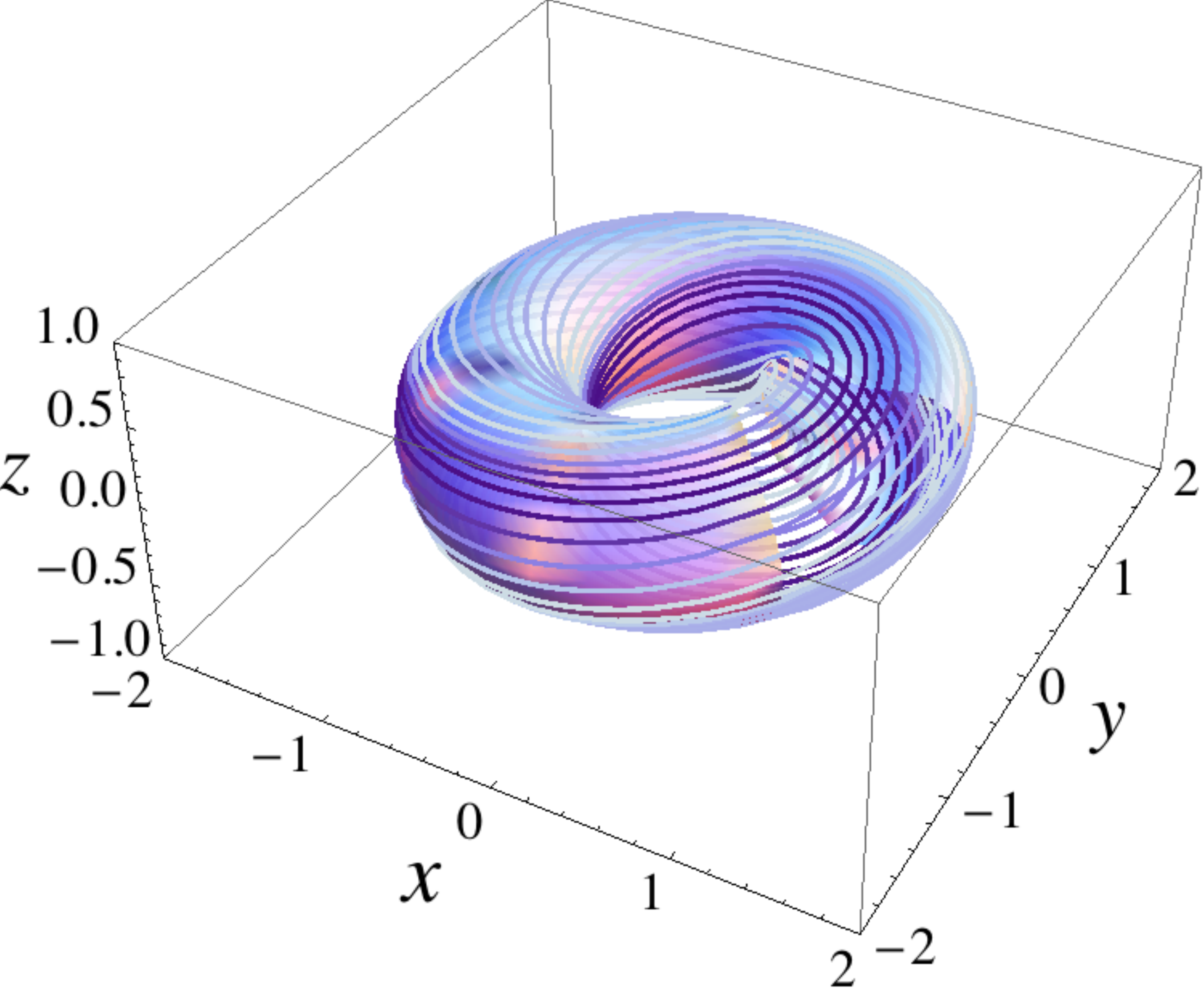}
	\end{tabular}
   \caption{Illustration of the graph $I=\rho(\theta,t)$ corresponding to the pseudo-orbits in Fig.~\ref{fig:psTorusConfig} plotted using \emph{Mathematica}'s \texttt{ParametricPlot3D} to show embeddings into $\mathbb{R}^3$ using two mappings: (a) Left: A simple Cartesian representation $x = t$, $y = \theta$, $z = I$. (b) Right: A 3D polar-like representation $x = [1 + \rho(\theta,t) \cos\theta] \cos t$, $[1 + \rho(\theta,t) \cos\theta] \sin t$, $z = \rho(\theta,t) \sin\theta$, analogous to an almost-invariant torus in magnetic-field-line flow.}
   \label{fig:psTorusEmbeddings}
\end{figure}

For simplicity (but see Remark~\ref{rem:GraphProp} below) assume the configuration-space paths do not intersect in the $\theta,t$ covering space, so the map $\vartheta: \theta_0 \mapsto \theta$ is a diffeomorphism for any value of $t$. As $\vartheta(t|\theta_0)$ increases monotonically with $\theta_0$, when $t = 0$, this remains true for all time:
\begin{equation}
	\frac{\partial\vartheta}{\partial\theta_0} > 0\; \forall\; t \;.
	\label{eq:nonintersect}
\end{equation}
Also, $I$ is a single-valued function of $\theta_0$ and $t$ and we can invert $\vartheta$ to make $I$ a graph over $\theta$ and $t$. That is, we can find a single-valued function $\rho$ such that $I = \rho(\theta,t)$. This is illustrated in Fig.~\ref{fig:psTorusEmbeddings}.

Equations~(\ref{eq:torusfoliation}) clearly show that foliation by a given pseudo-orbit family is sufficient to specify a torus. \emph{Conversely}, specification of an arbitrary trial torus $\mathcal{T} \in \Gamma_3$ implies a natural vector field in the tangent space at each point on the torus that locally defines a family of pseudo-orbits via a dynamical system. To see this, define $\mathcal{T}$ via a graph (for simplicity, but see Remark~\ref{rem:GraphProp}) of $I$ over the $\theta,t$ covering space $\mathbb{R}\times\mathbb{R}$	
\begin{equation}
	        \mathcal{T}[\rho]: I = \rho(\theta,t) \quad \forall  \:\theta,t \in \mathbb{R}\;,
	\label{eq:Tdef}
\end{equation}
where $\rho$ is a $2\pi$-periodic function of $\theta$ and $t$. Combined with  Eqs.~(\ref{eq:HamThdot}) [implied by the constraint Eq.~(\ref{eq:AgradHamThdot})] and (\ref{eq:Hamtdot}) this implies the following two-dimensional vector field
	\begin{equation}
	        [\dot\theta,\:\dot t\,](\theta,t)  \equiv   \left[H_I(\rho(\theta,t),\theta,t),\: 1\,\right] \;.
	\label{eq:Tthetaeqn}
	\end{equation}

\begin{remark}[Graph property]
	\label{rem:GraphProp}
The assumption in Eq.~(\ref{eq:Tdef}) that $\rho$ is a graph, i.e. single valued, is convenient but not strictly necessary. Following Dewar and Meiss \cite{Dewar_Meiss_92} and Dewar and Khorev \cite{Dewar_Khorev_95} we could relax this assumption by semi-conjugating $\theta$ to a linearly time-dependent family of rotations: $\theta = \vartheta(\Theta_0 + \omega t)$, which also puts the case of irrational $\omega$ on the same footing as the rational case. For simplicity we shall not pursue this approach further in the current paper, but it is sometimes found to be necessary for QFMin tori, as they are not always graphs over the angle coordinates $\theta$ and $t$. Ghost tori on the other hand, have, at least in the case of area-preserving twist maps, been proven to possess the graph property \cite{Gole_01}. 
\end{remark}

Eliminating $\tau$ in favor of $t$ yields a nonlinear, first-order o.d.e. defining a family of solutions $\theta = \vartheta(t|\theta_0)$, parametrized by the initial values $\theta_0$, 
	\begin{equation}
	        \d_t\vartheta(t|\theta_0)  =   H_I(\rho(\vartheta(t|\theta_0), t),\,\vartheta(t|\theta_0), t) \;.
	\label{eq:Tvarthetaeqn}
	\end{equation}
Thus we can define a phase-space pseudo-orbit $\theta = \vartheta(t|\theta_0),\; I =\mathcal{I}(t|\theta_0) \equiv \rho(\vartheta(t|\theta_0),t)$ passing through each point of $\mathcal{T}$.	
\begin{remark}[Condition for foliation]
	\label{rem:FoliationCond}
Integrating Eq.~(\ref{eq:Tvarthetaeqn}) once around the torus defines the circle map $\theta_0 \mapsto  \vartheta(2\pi|\theta_0)$.
Circle maps with rational rotation number generically have a singular invariant measure because of the phenomenon of Arnold tongues \cite{Arrowsmith_Place_91}. Thus the pseudo-orbits on an \emph{arbitrary} torus do \emph{not} in general smoothly foliate it. However, the Euler--Lagrange equation that will be derived in Sec.~\ref{sec:HamQFMin} for a QFMin torus generates pseudo-orbits that foliate of the torus smoothly so no inconsistency is generated by assuming such foliation from the outset.
\end{remark}

\subsection{Ghost tori}
\label{sec:ghost}
As illustrated in Fig.~{\ref{fig:SInteg}}, when an invariant torus with rational rotation number is destroyed by a perturbation, two true orbits with the same rotation number survive: an ``X-point'' orbit that is a minimum of $S$, and an ``O-point'' orbit that is a minimax (saddle) point of $S$. The ghost-curve strategy is to join these two periodic orbits by a family of pseudo-orbits, labeled by a continuous parameter $T(\theta_0)$, generated by flowing down the action gradient from minimax orbits to minimizing orbits. Below we describe this in configuration space and phase space.

\subsubsection{Ghost torus---Lagrangian approach}
\label{sec:LagGhost}

In configuration space the gradient flow equation is
\begin{equation}
	\frac{\D\vartheta}{\D T} \equiv \frac{1}{T'(\theta_0)}\frac{\partial\vartheta}{\partial\theta_0} =  -\frac{\delta{S}}{\delta\theta} \;.
	\label{eq:thetaGhostEqLag}
\end{equation}
This flow evolves a function, $\vartheta$, defined on the infinity of points on the interval $t \in [0,2\pi q]$. Furthermore, values at different $t$ are coupled because $\delta{S}/\delta\theta$, defined in Eq.~(\ref{eq:actiongrad}), involves first and second time derivatives of $\vartheta$. The problem is thus infinite-dimensional, but a discretized approximation can be solved numerically in a straightforward fashion \cite{Hudson_Dewar_09}.

The function $T(\theta_0)$ is not known \emph{a priori} but, once $\vartheta$ is found as a function of $T$, it may be calculated by integrating
\begin{equation}
	T'(\theta_0) =  -\left(\frac{\delta{S}}{\delta\theta}\right)_{t = 0}^{-1} \;,
	\label{eq:TGhostDef}
\end{equation}
which follows from Eq.~(\ref{eq:thetaGhostEqLag}) when it is recognized that $\partial\vartheta(0|\theta_0)/\partial\theta_0 \equiv 1$.
Because $\delta{S}/\delta\theta = 0$ at the minimizing and minimax periodic orbits, $T$ varies between $\mp\infty$. 

\subsubsection{Ghost torus---Hamiltonian approach}
\label{sec:HamGhost}

In phase space, pseudo-orbits are defined as in Def.~(\ref{def:Lpsorb}) and we thus need to evolve both $\vartheta$ and $\mathcal{I}$ with respect to $T$. The former is still evolved by the action-gradient flow
\begin{equation}
	\frac{\D\vartheta}{\D T} =  -\frac{\delta{S}_{\rm ph}}{\delta\theta} \;,
	\label{eq:thetaGhostEqHam}
\end{equation}
but, to maintain equivalence with the Lagrangian formulation, the constraint Eq.~(\ref {eq:AgradHamThdot}) must be used in the form $\D_T(\delta S_{\rm ph}/\delta I) \equiv \D_T(\d_t\vartheta - H_I) = 0$, where $\d_t \equiv \d/\d t$ and $\D_T \equiv \D/\D T$. As $t$ and $T$ are independent variables, we can interchange the order of $\d_t$ and $\D_T$. Using Eq.~(\ref{eq:thetaGhostEqHam}), we find the evolution equation for $\mathcal I$
\begin{equation}
	\frac{\D\mathcal{I}}{\D T} =  \frac{1}{H_{II}}
		\left(
		  H_{I\theta} - \frac{\d}{\d t}
		\right)\frac{\delta{S_{\rm ph}}}{\delta\theta}  \;.
	\label{eq:IGhostEq}
\end{equation}
Interestingly, the right-hand side of Eq.~(\ref{eq:IGhostEq}) is seen from Eq.~(\ref{eq:POrbphQFEL1}) to be the Lagrange multiplier $\lambda$ arising from the AGMin construction. However, Eq.~(\ref{eq:POrbphQFEL2}) is not consistent with Eq.~(\ref{eq:thetaGhostEqHam}) so ghost pseudo-orbits are not a subset of AGMin orbits.

\subsection{QFMin tori}
\label{sec:QFMinTori}

We now define functionals $\varphi_1$ and $\varphi_2$ as surface integrals of powers of the action gradient over $\mathcal{T}$
\begin{eqnarray}
	\varphi_1 & \equiv &
	\int_0^{2\pi}\!\!\!\!\int_0^{2\pi}\frac{\delta S}{\delta\theta} \, \d\theta\d t \;,
		\label{eq:phi1}\\
	\varphi_2 & \equiv &
	\frac{1}{2} \iint\limits_0^{\quad\: 2\pi}\left(\frac{\delta S}{\delta\theta}\right)^2 \d\theta \d t \;.
		\label{eq:phi2}
\end{eqnarray}

It can be shown [see Eq.~(\ref{eq:Tactiongrad}) below] that the surface integral linear in the action gradient vanishes identically, i.e. $\varphi_1 \equiv 0$, which supports the interpretation of $\varphi_1$ as the net \emph{flux} of extended phase space volume across $\mathcal{I}$. (In the case of magnetic fields this is the net magnetic flux \cite{Hudson_Dewar_96}, which is zero because of the absence of magnetic monopoles.)

Thus we must go to the ``quadratic flux'' $\varphi_2 \geq 0$ before having a nontrivial measure of the invariance of $\mathcal{T}$ under the dynamics  (equality to zero applying when $\delta S/\delta\theta = 0$ on $\mathcal{T}$, i.e. if and only if it is an invariant torus). The quadratic flux is the analogue of the objective functional Eq.~(\ref{eq:POrbQF}) for action-gradient-minimizing (AGMin) pseudo-orbits, but with the $L^2$ norm defined on tori rather than paths. Analogously, we define a quadratic-flux minimizing (QFMin) \emph{torus} as follows.
\begin{definition}[QFMin torus]
A QFMin torus $\cal T$ is one that minimizes $\varphi_2$.
\end{definition}

\subsubsection{QFMin torus---Lagrangian approach}
\label{sec:LagQFMin}

In the Lagrangian approach we define the trial torus $\mathcal{T}[\vartheta]$ in terms of arbitrarily variable pseudo-orbits using Eq.~(\ref{eq:torusfoliation}), which requires Eq.~(\ref{eq:phi2}) to be modified due to the change of variable $\theta \mapsto \theta_0$, giving the alternative Lagrangian definition 
\begin{equation}
  \begin{split}
	\varphi_2 & \equiv \frac{1}{2}\int_0^{2\pi q}\!\!\!\d t
					\!\! \int_{\theta_0^{(0)}}^{\theta_0^{(1)}}\!\!\!\d\theta_0 \,
				\frac{\partial\vartheta}{\partial\theta_0}\left(\frac{\delta S}{\delta\theta}\right)^2\\
	&  = \frac{1}{2}\int_0^{2\pi} \!\!\!\d t\sum_{i=0}^{q-1} 
								 \int_{\theta_0^{(i)}}^{\theta_0^{(i+1)}}\!\!\!\d\theta_0
	\frac{\partial\vartheta}{\partial\theta_0}\left(\frac{\delta S}{\delta\theta}\right)^2 \\
	&  = \frac{1}{2} \iint\limits_0^{\quad\: 2\pi}
		\left(\frac{\delta S}{\delta\theta}\right)^2 \!\frac{\partial\vartheta}{\partial\theta_0}\,
		\d\theta_0 \d t \;,
  \end{split}
	\label{eq:phi2alt}
\end{equation}
the manipulations required to arrive at the latter form being  based on the assumption that the integrand is periodic in $\theta_0$ and $t$, and the endpoints $\theta_0^{(i)}$ being members of the ordered set of return points of the orbit starting at $\theta_0 = 0$ defined at the beginning of Sec.~\ref{sec:pstor}.

Varying Eq.~(\ref{eq:phi2alt}) and integrating by parts to remove the $\theta_0$ derivative from $\delta\vartheta$, we find the first variation
\begin{equation}
	\delta\varphi_2[\vartheta_{p,q}]  =  \!\! \iint\limits_0^{\quad\: 2\pi} \frac{\delta S}{\delta\theta}
	\left[
		\frac{\partial\vartheta}{\partial\theta_0}\delta\left(\frac{\delta S}{\delta\theta}\right) 
		- \delta\vartheta\frac{\partial}{\partial\theta_0}\left(\frac{\delta S}{\delta\theta}\right) 
	\right]\d\theta_0 \d t \;.
	\label{eq:deltaphi2Lag}
\end{equation}

With $\theta_0$ as an independent variable, the total time derivative along a pseudo-orbit, $\d_t \equiv \d/\d t$, is $\partial/\partial t$, which  commutes with $\partial/\partial\theta_0$. Thus we can use Eq.~(\ref{eq:POrbdeltaactiongrad}) to evaluate both terms in Eq.~(\ref{eq:deltaphi2Lag}). The terms not involving $\d_t$ cancel, leaving
\begin{equation}
\begin{split}
	\delta\varphi_2[\vartheta_{p,q}]  & = \!\!\iint\limits_0^{\quad\: 2\pi}\d\theta_0 \d t \,
	\frac{\delta S}{\delta\theta}
	\left\{
		\frac{\partial\vartheta}{\partial\theta_0}
		\d_t	\left[
			\left(\frac{\partial L_{\dot\theta\dot\theta}}{\partial t}\right)\delta\vartheta
				- \d_t\left(L_{\dot\theta\dot\theta}\delta\vartheta\right)
			\right] \right. \\  & \left. \hspace{2.5cm}
			\mbox{} - \delta\vartheta\,
			\d_t\left[
				\left(\frac{\partial L_{\dot\theta\dot\theta}}{\partial t}\right)\frac{\partial\vartheta}{\partial\theta_0}
				- \d_t\left(L_{\dot\theta\dot\theta}\frac{\partial\vartheta}{\partial\theta_0}\right)
			\right]
	 \right\} \\
	 & =  - \iint\limits_0^{\quad\: 2\pi} \d\theta_0 \d t \left(\frac{\partial\vartheta}{\partial\theta_0}\right)^{-1}\,
	 \delta\vartheta\,
	 \d_t\,\left[
	 	L_{\dot\theta\dot\theta}\left(\frac{\partial\vartheta}{\partial\theta_0}\right)^2
			\d_t\frac{\delta S}{\delta\theta}
	 \right]
	 \;,
	 \label{eq:deltaphi2LagSimp}
\end{split}
\end{equation}
where the second form is obtained after integration by parts in $t$, leading to multiple cancellations, and we assumed Eq.~(\ref{eq:nonintersect}) to justify dividing by $\partial\vartheta/\partial\theta_0$. 

Equating $\delta\varphi_2$ to zero for all $\delta\vartheta$ yields the Euler--Lagrange equation for the periodic pseudo-orbits making up a QFMin torus
\begin{equation}
	\frac{\d}{\d t}\left[
	 	L_{\dot\theta\dot\theta}\left(\frac{\partial\vartheta}{\partial\theta_0}\right)^2
			\frac{\d}{\d t}\frac{\delta S}{\delta\theta}
	 \right] = 0 \;,
	\label{eq:QFMinELLag}
\end{equation}
where we have reverted to the equivalent notation $\d/\d t$ for the time derivative along a given pseudo-orbit (i.e. $\theta_0$ fixed). Remarkably, and unlike the Euler--Lagrange equations for AGMin pseudo-orbits, this equation may be analytically integrated to give the action gradient on the QFMin-torus pseudo-orbits,
\begin{equation}
	\frac{\delta S}{\delta\theta} = \nu 
	+ \mu\int_0^t \left[L_{\dot\theta\dot\theta}\left(\frac{\partial\vartheta}{\partial\theta_0}\right)^2\right]_{t\mapsto t'}^{-1}\d t' \;,
	\label{eq:QFMinThetaLag}
\end{equation}
where $\mu(\theta_0)$ and $\nu(\theta_0)$ are constants of integration.
Assuming the twist condition, Eq.~(\ref{eq:ShearCond}), $L_{\dot\theta\dot\theta} = 1/H_{II} \neq 0$, so the second term is monotonically secular in $t$ unless $\mu = 0$. However  $\delta S/\delta\theta$ is evaluated on a periodic pseudo-orbit so it must be periodic, implying $\mu = 0$ is the \emph{only} allowed choice. The meaning of the remaining constant, $\nu$, will be discussed further below after we derive the same equation using a Hamiltonian approach.

\subsubsection{QFMin torus---Hamiltonian approach}
\label{sec:HamQFMin}

Consider a trial torus $\mathcal{T}[\rho]$ as defined in Eq.~(\ref{eq:Tdef}). Using Eqs.~(\ref{eq:Tdef}) and (\ref{eq:Tthetaeqn}) we define a Hamiltonian vector field on the $\theta,t$ plane
\begin{equation}
        [\dot\theta,\:\dot I\,](\theta,t)
        	\equiv
        [H_I,\: \rho_t + H_I\, \rho_{\theta}] \;,
\label{eq:THamvecfld}
\end{equation}
where $H_I$ denotes $H_I(\rho(\theta,t),\theta,t)$, and $\rho_t$ and $\rho_{\theta}$ denote the partial derivatives of $\rho(\theta,t)$ with respect to $t$ and $\theta$.

Using Eq.~(\ref{eq:THamvecfld}) to eliminate $\dot\theta$ from Eq.~(\ref{eq:phactiongradI}) we have $\delta S_{\rm ph}/\delta I \equiv 0$ and, eliminating $\dot I$ from Eq.~(\ref{eq:phactiongradtheta}), $\delta S/\delta\theta \equiv \delta S_{\rm ph}/\delta\theta$ becomes a scalar field on the $\theta,t$ plane
\begin{eqnarray}
	\frac{\delta S}{\delta\theta}
	& =&  -(\rho_t + H_I\, \rho_{\theta}) - H_{\theta} \nonumber\\
	& \equiv&  -\rho_t - \eth_{\theta} H \;,
	\label{eq:Tactiongrad}
\end{eqnarray}
where $H_{\theta}$ denotes $H_{\theta}(\rho(\theta,t),\theta,t)$, whereas $\eth_{\theta} H$ denotes the total $\theta$ derivative, $H_{\theta}+ H_I\,\rho_{\theta}$.

The QFMin approach to defining almost-invariant tori is to minimize $\varphi_2$ over arbitrary deformations of $\mathcal{T}$.  Thus we consider infinitesimal deformations generated by variations $\delta\rho(\theta,t)$. From Eq.~(\ref{eq:Tactiongrad}), 
\begin{equation}
	\delta\frac{\delta S}{\delta\theta}
	= -\delta\rho_t - \eth_{\theta} (H_I\delta\rho) \;.
	\label{eq:Tactiongradvar}
\end{equation}
Using this result in the first variation of Eq.~(\ref{eq:phi2}) and integrating by parts we find
\begin{equation}
	\delta\varphi_2 = 
	\iint\limits_0^{\quad\: 2\pi}
	\delta\rho \,\d_t  \left(\frac{\delta S}{\delta\theta}\right) \d\theta \d t
	\label{eq:phi2var}
\end{equation}
where $\d_t$ denotes $\partial_t + H_I\eth_{\theta}$.

Equating $\delta\varphi_2$ to zero for all variations $\delta\rho$ gives the Hamiltonian analogue of Eq.~(\ref{eq:QFMinELLag}) 
\begin{equation}
	\d_t  \left(\frac{\delta S}{\delta\theta}\right)  = 0 \;.
	\label{eq:QFMinELHam}
\end{equation}
Integrating Eq.~(\ref{eq:QFMinELHam}) along the path defined by Eq.~(\ref{eq:Tvarthetaeqn}) gives Eq.~(\ref{eq:QFMinThetaLag}) with $\mu = 0$. Thus, using either the Hamiltonian or the less elegant Lagrangian approach, we are led to
\begin{definition}[QFMin pseudo-orbit]
	\label{def:TorusQFMinPath}
A QFMin pseudo-orbit is one satisfying the constraint Eq.~(\ref{eq:AgradHamThdot}) and the \emph{modified} Hamiltonian or Lagrangian equation of motion
\begin{equation}
	\frac{\delta S}{\delta\theta} = \nu \;,
	\label{eq:QFMinThetaHam}
\end{equation}
where $\nu(\theta_0)$ is constant along each pseudo-orbit.
\end{definition}
Then the desired QFMin torus is constructed by varying $\nu$ continuously over its range so as as to sweep out $\mathcal{T}$ by QFMin orbits foliating the surface. Note that $\nu = 0$ on the action-minimizing or -minimax orbits of an island chain, so these are automatically incorporated into $\mathcal{T}$ as in the ghost-orbit construction. 

For low-order periodic orbits in a system close to integrability, i.e. for which Eq.~(\ref{eq:PertH}) applies, $\nu = O(\epsilon)$ and a perturbation expansion can be used. Numerically, a QFMin torus with $\omega = p/q$ may easily be found \cite{Hudson_Dewar_96} by integrating Eq.~(\ref{eq:QFMinThetaHam}), with the constraint Eq.~(\ref{eq:AgradHamThdot}), given initial values $\theta_0$ and $I_0$ (or $\dot\theta_0$) on the Poincar{\'e} section $t = 0$ and a guess for $\nu$. Holding $\theta_0$ fixed, $\nu$ and $I_0$ are adjusted iteratively in a two-dimensional search until a $(p,q)$-pseudo-orbit satisfying the periodicity conditions Eqs.~(\ref{eq:pqperiodicity}) is found to the desired accuracy. As this procedure can be carried out for all $\theta_0$, the desired foliation of the torus is achieved.


\section{Conclusion}
\label{sec:Concln}

We have reviewed equivalent general Lagrangian and Hamiltonian action-based formulations for defining and calculating almost-invariant tori based on the concept of pseudo-orbit. In previous work \cite{Hudson_Dewar_09} we have found the QFMin approach to be easier to implement numerically, but the ghost surface approach is more satisfactory in that the graph property is preserved for high-order orbits and high nonlinearity (cf. Remark~\ref {rem:GraphProp}). 
Some open issues are:
\begin{itemize}
\item Can we unify the rational-$\omega$ QFMin and ghost surface approaches for arbitrary nonlinearity through an appropriate change of coordinates \cite{Dewar_Hudson_Gibson_10}?
\item If so, can irrational-$\omega$ unified tori be defined as the limit of a sequence of $p,q$-almost-invariant tori up to and beyond the breakup of the corresponding invariant torus? (In which case the quadratic flux could be used as an alternative to Greene's residue \cite{MacKay_92} for determining existence as well as a measure of transport through the cantorus beyond breakup.)
\item We can also define an ``AGMin'' pseudo-invariant torus by replacing the boundary conditions $\vartheta'(0) = \dot\theta_0$ and $\vartheta'(2\pi q) = \dot\theta_0$ in Eq.~(\ref{eq:pqperiodicityconds}) with $\vartheta'(0) = \vartheta'(2\pi q)$ and $\vartheta''(0) = \vartheta''(2\pi q)$, so that $\dot\theta_0$ is selected to make the pseudo-orbits analytic for all $t$ and $\theta_0$. What are the properties of such a torus?
\item Can we usefully widen the allowed class of phase-space pseudo-orbits by relaxing the constraint Eq.~(\ref{eq:AgradHamThdot})?
\item Can we further generalize the phase-space pseudo-orbits by using noncanonical approaches?
\end{itemize}

\section*{Acknowledgement}

One of the authors (RLD) thanks the hospitality of the Nonlinear Dynamics Group at the CNRS Centre de Physique Th\'eorique, Luminy, Marseille, France, where some of this work was performed and discussed, including useful conversations with Professor Philip Morrison. Author SRH acknowledges support from U.S. Department of Energy Contract No. DE-AC02-09CH11466 and Grant No. DE-FG02-99ER54546.

\bibliographystyle{elsarticle-num} 
\bibliography{RLDBibDeskPapers}

\end{document}